\documentclass[prc,twocolumn,aps,nofootinbib, superscriptaddress]{revtex4}
\usepackage{graphicx}

\begin{document}
\date{\today}
\title{{\bf Stochastic Schroedinger equation from optimal observable evolution }}

\author{Denis Lacroix}
\affiliation{National Superconducting Cyclotron Laboratory, Michigan State University, 
East Lansing, Michigan 48824, USA }
\affiliation{
Laboratoire de Physique Corpusculaire,
ENSICAEN and Universit\'e de Caen,IN2P3-CNRS,
Blvd du Mar\'{e}chal Juin,\\
14050 Caen, France}

\begin{abstract}
In this article, we consider a
set of trial wave-functions denoted by $\left| Q \right>$ and an associated set of operators $A_\alpha$ 
which generate transformations connecting those trial states.
Using variational principles, we show that we can always obtain a quantum Monte-Carlo method where 
the quantum evolution of a system is replaced by 
jumps between density matrices of the form $D = \left| Q_a \right>\left< Q_b \right|$,
and where the average evolutions of the moments of $A_\alpha $ up to a given order $k$, i.e. 
$\left< A_{\alpha_1} \right>$, $\left< A_{\alpha_1} A_{\alpha_2} \right>$,...,
$\left< A_{\alpha_1} \cdots A_{\alpha_k} \right>$, are constrained to follow the exact Ehrenfest evolution at 
each time step along each stochastic trajectory. Then, a set of more and more elaborated stochastic approximations 
of a quantum problem is obtained which approach the exact solution 
when more and more constraints are imposed, i.e. when $k$ increases. 
The Monte-Carlo process might even become exact if the Hamiltonian $H$ applied on the trial state can be 
written as a polynomial of $A_\alpha$. The formalism makes a natural connection between quantum 
jumps in Hilbert space and phase-space dynamics. We show that the derivation of stochastic Schroedinger 
equations can 
be greatly simplified by taking advantage of the existence of this hierarchy of approximations and its 
connection to the Ehrenfest theorem. Several examples are illustrated: the free wave-packet expansion, 
the Kerr oscillator, a generalized version of the Kerr oscillator, as well as interacting bosons or 
fermions.    
\end{abstract}

\maketitle

\section{Introduction}

Starting from the pioneering works of Feynman on path integrals \cite{Fey65}, 
large efforts have been devoted to the possibility to replace a quantum problem by a 
stochastic process \cite{Cep95,Neg88,Koo97,Ple98,Bre02,Sto02}. One of the greatest interest
of stochastic formulations can be summarized as follow: 
let us consider a quantum system where the number of degrees of freedom 
to follow is too high to authorize an exact treatment. 
In some cases, it is possible to recover the exact solution by averaging over 
an ensemble of trajectories where the number of 
degrees of freedom to consider along each path is significantly 
reduced. Then the exact problem can be recast into a set of problems that can be carried out. Such a stochastic 
formulation is for instance at the heart of quantum Monte-Carlo techniques 
in particular when considering interacting systems \cite{Cep95,Neg88,Koo97}.

Essentially two strategies exist to introduce stochastic formulations.
The first technique consists of selecting a limited number of degrees of freedom 
associated to specific observable, which contains 
already a large amount of information on the system. Then, the system evolution 
is recovered by averaging over stochastic evolution of these observable.  
This technique will be called here as
'Phase-Space technique' \cite{Gar00}. New developments along this line have been proposed 
to treat interacting fermionic systems \cite{Cor04}. 

Alternatively, several works have shown that one can take advantage of the Stochastic Schroedinger 
Equation (SSE) approach
\cite{Dio85,Gis84,Ghi90,Dal92,Dum92,Gis92,Car93,Dio94,Ima94,Bre95,Rig96,Cas96,Ple98,
Gar00,Bre02,Sto02} 
to simulate exactly quantum systems \cite{Car01,Jul02,Jul04,Bre04,Lac05-2}. 
In that case, a specific class of trial wave-functions is selected and the exact Schroedinger 
equation is mapped into a stochastic process between trial 
states. Again, trial wave-functions are chosen to be much simpler than the exact 
wave-function, although complex enough to incorporate already some important features of the problem. 
SSE have generally an equivalent stochastic evolution in phase-space: 
for instance, the stochastic process proposed in ref. \cite{Jul02}
can be equivalently replaced by a stochastic process for the one-body density evolution \cite{Lac05}
associated with N-body densities $D = \left| \Phi_1 \right> \left< \Phi_2 \right|$ where both states 
corresponds to trial wave-functions, i.e. Hartree-Fock states. 

In this work, we would like to address more systematically the 
possibility to obtain SSE for a given quantum problem using the phase-space 
evolution.
Following ref. \cite{Fel00}, we consider trial state vectors denoted 
by $\left| Q \right>$, where $Q=\left\{ q_\alpha \right\}_{\alpha =1,N}$ (with $N$ eventually infinite)
correspond to a set of variables which completely determines the state. The ensemble of states 
obtained by taking all possible values of $Q$ is denoted by ${\cal S}_Q$.
The second important hypothesis is that we assume the existence of an ensemble 
of operators $\left\{ A_\alpha \right\}_{\alpha =1,N}$, that generate local transformations 
between the state of ${\cal S}_Q$, i.e. one state $\left| Q \right>$ 
transforms into another state $\left| Q + \delta Q\right>$ of ${\cal S}_Q$ 
through the transformation 
\begin{eqnarray}
\left| Q + \delta Q \right> =  e^{\sum_{\alpha} \delta q_\alpha A_\alpha} \left| Q \right>.
\label{eq:eqa}
\end{eqnarray}
At the deterministic level, variational principles \cite{Bla86,Fel00} provide a systematic way to 
replace the exact problem by an approximate dynamics in restricted space ${\cal S}_Q$. 
A first example of extension of variational methods to obtain stochastic Schroedinger
equation has been given in ref. \cite{Wil03}. 

In this article we propose an alternative procedure to obtain SSE from variational principles.
Some aspects related 
to standard variational methods are first recalled. Schroedinger equations obtained in this way are intimately 
connected to the phase-space evolution and can be regarded as the best projected approximation for the dynamics 
within the subspace of observable $\left< A_\alpha  \right>$.
We show that variational techniques can be helpful to obtain stochastic formulations 
of a given problem which has a natural interpretation in terms of observable evolution.  
Starting from variational methods, we prove the existence of a hierarchy 
of approximations of a quantum problem using SSE methods where the expectation values of the moments 
of the $A_\alpha $ up to a given order $k$, i.e. 
$\left< A_{\alpha_1} \right>$, $\left< A_{\alpha_1} A_{\alpha_2} \right>$,...,
$\left< A_{\alpha_1} \cdots A_{\alpha_k} \right>$, follow in average the exact Ehrenfest evolution 
at each time step along each stochastic trajectory. 
Therefore, this generalization of the 
deterministic case can be regarded as the best approximation of the dynamics 
on an enlarged subspace of observable.

In the second part of this article, the existence of such 
a hierarchy and its connection to the Ehrenfest evolution are used  
to deduce SSE in several illustrative examples.

\section{General remarks on variational principles}
\label{sec:varia1}

Variational principles are powerful tools to provide approximate solutions 
for the static or dynamical properties of a system when few degrees of freedom 
are supposed to dominate\cite{Ker76,Kra81,Bla86,Dro86,Bro88,Fel00}. The selection of few 
relevant degrees of freedoms generally strongly reduces the complexity of the considered problem.
Before introducing stochastic mechanics, we consider standard variational principles 
and recall specific aspects which will be helpful 
in the following discussion. We assume that
the system is given at time $t_0$ by $\left| \Psi(t_0) \right>=\left| Q \right>$
where $\left| Q \right>$ belongs to ${\cal S}_Q$.
The wave-function, denoted by $\left| \Psi(t) \right>$, which evolves according to a given Hamiltonian $H$ 
has a priori no reason to remain in the subspace ${\cal S}_Q$. 
However, a good approximation of the dynamics 
restricted to wave-functions of ${\cal S}_Q$ is obtained by minimizing the 
action\cite{Bla86,Fel00}
\begin{eqnarray}
S = \int_{t_0}^{t_1} 
ds \left<  Q \right|  i\hbar \partial_t - H  \left|  Q  \right>,
\label{eq:varia}
\end{eqnarray} 
with respect to the variation of $Q$ and 
with the additional boundary condition $\left|  \delta \Phi(t_0)  \right> =0$ and 
$\left< \delta \Phi(t_1) \right|=0$. 
The variation with respect to $\left<  \delta Q \right|$ leads then to the equation
\begin{eqnarray}
\left<  \delta Q \right|  i\hbar \partial^{\triangleright}_t - H  \left|  Q  \right> = 0,
\end{eqnarray} 
while the variation with respect to $\left| \delta  Q \right>$ gives the hermitian conjugate 
equation 
\begin{eqnarray}
\left<   Q \right|  i\hbar \partial^{\triangleleft}_t + H  \left|  \delta Q  \right> = 0.
\end{eqnarray} 
Here $\partial^{\triangleleft}_t $ and $\partial^{\triangleright}_t$ means that the derivative is acting 
respectively to the left and right hand side. 

\subsection{observable evolution}

Due to relation (\ref{eq:eqa}), the above Schroedinger equation corresponds to specific evolutions of the 
expectation values $\left< A_\alpha \right>$ of the operators $A_\alpha $.  
Using $\left| \delta Q \right> = 
\sum_\alpha \delta q_\alpha A_\alpha \left| Q  \right>$, variations with respect to
$\left|  \delta Q \right>$ are replaced by a set of independent variations 
$\left\{\delta q_\alpha\right\}_{\alpha =1,N}$, 
leading to: 
\begin{eqnarray}
i \hbar \left< d Q \left| A_\alpha \right| Q \right> =-
dt \left< Q \left| H A_\alpha \right| Q \right>,
\end{eqnarray}
while variation with respect to $\delta q^*_\alpha$ gives
\begin{eqnarray}
i \hbar \left< Q \left| A_\alpha \right| d Q \right> =dt \left< Q 
\left| A_\alpha  H\right| Q \right>. 
\label{eq:ada}
\end{eqnarray} 
Combining the two equations gives
\begin{eqnarray}
i \hbar \frac{d\left< A_\alpha \right>}{dt} &=& \left< \left[ A_\alpha , H \right] \right> ,
\end{eqnarray}   
which is nothing but the exact Ehrenfest equation of motion for the set of operators $A_\alpha$.
Therefore, starting from a density $D(t_0) = 
\left| Q \right>\left< Q \right|$, 
for one time step the evolution of $<A_\alpha>$ corresponds 
to the exact evolution although the state is constrained to remain in the ${\cal S}_Q$ space. 
This property is however a priori only true for observable which are linear
combination of $A_\alpha$. 
     
\subsection{Links with effective Hamiltonian and projected dynamics}

The evolution deduced from the variational method 
can equivalently be interpreted as an effective Hamiltonian dynamics.
Indeed, we have 
\begin{eqnarray}
\left| dQ \right> = 
\sum_\alpha d q_\alpha A_\alpha \left| Q \right> \equiv \frac{dt}{i \hbar} H_{1} \left| Q \right>,
\label{eq:dqq} 
\end{eqnarray}  
where we have introduced the effective Hamiltonian $H_1$. This Hamiltonian differs from $H$ unless $H$ applied 
on the trial wave-function is itself a linear combination of $A_\alpha$. 
The expression of $H_1$ can be obtained using the evolution of $dq_\alpha$. From eq. (\ref{eq:ada}) we obtain
\begin{eqnarray}
i \hbar \sum_\alpha dq_\alpha \left<   A_\beta A_\alpha \right> = dt \left<  A_\beta H  \right>.
\label{eq:aaab}
\end{eqnarray}
Writing 
$\left< Q \left| A_\beta A_\alpha \right| Q \right>=M_{\beta \alpha}$, 
$d q_\alpha$ can formally be obtained by inverting the last expression into
\begin{eqnarray}
dq_\alpha = \frac{dt}{i \hbar} \sum_\beta M^{-1}_{  \alpha \beta } \left< Q \left| A_\beta  H \right| Q \right>.
\end{eqnarray}
Plugging last equation into (\ref{eq:dqq}) leads to $H_1 = {\cal P}_1 H$, where ${\cal P}_1$ is defined as 
\begin{eqnarray}
{\cal P}_1 = \sum_{\alpha \beta} A_\alpha \left| Q \right>M^{-1}_{  \alpha \beta} 
\left< Q \right|A_\beta ,
\end{eqnarray}
and plays the role of a "projector" onto the subspace of operators $A_\alpha$. 
In particular, for any observable 
$B$, the operator $B' = (1-{\cal P}_1)B$ fulfills $\left< A_\alpha B' \right>=0$. 
Similarly, if we denote by $B'' = B(1-{\cal P}_1)$, we have 
$\left< B''A_\alpha  \right>=0$. 
Writing the full Hamiltonian as 
\begin{eqnarray}
H &=& {\cal P}_1 H + (1-{\cal P}_1) H, \nonumber
\end{eqnarray} 
we see that the effective evolution of $\left| dQ \right>$ given by eq. (\ref{eq:dqq}) where $H_1$
is obtained from the minimization of the action can be interpreted as the approximate evolution when 
the part of the Hamiltonian that do not contribute to the $\left< A_\alpha \right>$ is neglected. 
As a matter of fact, the second part is responsible from the deviation of the projected dynamics 
from the exact one. In summary, we have recalled here that the variational dynamics 
can be regarded as a projected dynamics into the subspace of relevant observable \cite{Bal95,Bal99}.     
In the following, we show that aspects of variational principles presented here can be adapted 
to obtain Monte-Carlo methods. 

\section{Stochastic quantum mechanics from variational principles}

Several recent studies have shown that the exact evolution of complex quantum systems can be 
replaced by a set of stochastic evolutions of simpler trial wave-functions.
In this context, at a given intermediate time, the exact 
density is replaced by the average over "densities" of the form \cite{Car01,Jul02,Jul04,Bre04,Lac05-2} 
\begin{eqnarray}
D(t) = \left| Q_a \right> \left< Q_b \right|,
\label{eq:dab}
\end{eqnarray}  
where both states belongs to ${\cal S}_Q$. In all cases, the main advantage of these reformulations
was to prove that the stochastic methods can be performed 
imposing that both states remain in ${\cal S}_Q$. As in previous section, this condition implies that 
we restrict variations of each state to 
\begin{eqnarray}
\left| Q_a + \delta Q_a \right>  &=& e^{\sum_\alpha \delta q^{[a]}_\alpha A_\alpha}  \left| Q_a \right>, \\
\left| Q_b + \delta Q_b \right>  &=& e^{\sum_\alpha \delta q^{[b]}_\alpha A_\alpha}  \left| Q_b \right>,
\label{eq:statevaria}
\end{eqnarray}
where now $\delta q^{[a]}_\alpha$ and $\delta q^{[b]}_\alpha$ may also contain a stochastic part.

The aim of the present section is to show that, given a class of trial states, a hierarchy of 
Monte-Carlo formulations of a quantum problem can be systematically obtained 
using variational methods. 
The description of the system can be gradually improved by introducing a set of noises, written as 
\begin{eqnarray}
\left\{
\begin{array} {c}
\delta q^{[a]}_\alpha = \delta q^a_\alpha + \delta \xi^{[2]}_\alpha + \delta \xi^{[3]}_\alpha+ \cdots  \\  
\delta {q^{[b]}_\alpha}^* = \delta {q^b_\alpha}^* + \delta \eta^{[2]}_\alpha + \delta \eta^{[3]}_\alpha+ \cdots   
\end{array}
\right.
\label{eq:dqdq}
\end{eqnarray}
where the second, third... terms represent stochastic variables added on top of the deterministic 
contribution. Those are optimized to not only insure that the average evolution of $\left< A_\alpha \right>$ 
matches
the exact evolution
at each time step but also that the average evolutions of higher moments 
$\left< A_\alpha A_\beta \right>$, $\left< A_\alpha A_\beta A_\gamma \right>$,... follow the exact Ehrenfest 
dynamics.  
       
\subsection{Step 1: deterministic evolution}
\label{sec:step1}

Assuming first that stochastic contributions $\xi^{[i]}_\alpha$ and $\eta^{[i]}_\alpha$ are
 neglected in eq. (\ref{eq:dqdq}), 
we show how variational principles described previously can be used for mixed densities given by eq. (\ref{eq:dab}).
It is worth noticing that variational principle have also been proposed to 
estimate transition amplitudes \cite{Bla86} (see also discussion in \cite{Bal85}). 
In that case, different states are used in the left and right hand side of the action. 
This situation is similar to the case we are considering. 
We are interested here
in the short time evolution of the system, therefore we 
disregard the time integral in equation (\ref{eq:varia}) and consider directly the action    
\begin{eqnarray}
S = Tr\left( \left\{ i \hbar \partial^\triangleright_t - i \hbar \partial^\triangleleft_t - H  \right\} D \right), 
\end{eqnarray}
where $Tr(.)$ stands for the usual trace. 
Starting from the above action, one can generalize the different aspects discussed in section \ref{sec:varia1} to 
the case of densities formed with couples of trial states. For instance, the minimization with respect to the variations
$\left< \delta Q_b \right|$ and $\left| \delta Q_a \right>$ leads to the two conditions 
\begin{eqnarray}
\left\{
\begin{array} {c}
i \hbar \left< Q_b \left| A_\alpha \right| d Q_a \right> = \left< Q_b \left| A_\alpha  H \right| Q_a \right>, \\
\\
i \hbar \left< d Q_b \left| A_\alpha \right| Q_a \right> = \left< Q_b \left|  H A_\alpha \right| Q_a \right>,
\end{array}
\right.
\end{eqnarray}
from which we deduce that
\begin{eqnarray}
i \hbar \frac{d}{dt} \left< A_\alpha  \right> = \left< \left[ A_{\alpha}, H \right] \right>,
\end{eqnarray}
where $\left< A_\alpha  \right> = \left< Q_b \left| A_\alpha \right| Q_a \right>$. Therefore, the minimization of the action again insures that the exact Ehrenfest evolution is followed by the $A_\alpha$ observable
over one time step. Similarly, the evolution of both $\left| Q_a \right>$ and $\left| Q_b \right>$ are given by  
\begin{eqnarray}
\left\{
\begin{array} {ccc}
\left| dQ_a \right> &=& \sum_\alpha d q^{a}_\alpha A_\alpha \left| Q_a \right>=
\frac{dt}{i \hbar} {\cal P}_1 H\left| Q_a \right> \nonumber \\
\\
\left< dQ_b \right| &=& \left< Q_b \right|\sum_\alpha d {q^{b}_\alpha}^*A_\alpha = -\frac{dt}{i \hbar} 
\left< Q_b \right|H {\cal P}_1 \nonumber
\end{array}
\right.
\end{eqnarray}  
where ${\cal P}_1$ now reads
\begin{eqnarray}
{\cal P}_1 = \sum_{\alpha \beta} A_\alpha \left| Q_a \right>M^{-1}_{  \alpha \beta} 
\left< Q_b \right|A_\beta.
\end{eqnarray}
In opposite to previous section, ${\cal P}_1$ cannot be interpreted as a projector 
onto the space of observable due to the fact that $M_{\alpha \beta } = \left< Q_b\left| A_\alpha A_\beta \right| 
Q_a\right>$ is not anymore a metric for that space. 
However, we still have the property that the 
total Hamiltonian can be split into two parts, one which corresponds to the effective dynamic 
solution of the minimization, and one which is neglected, which identifies the irrelevant part
 for the $\left< A_\alpha \right>$ evolutions.  
It is worth noticing that the variational approximation presented here 
for mixed densities has been already used without justification in 
ref. \cite{Lac05-2} to optimize stochastic trajectories. 

\subsection{Step 2 : Introduction of gaussian stochastic processes}

In this section, we show 
that the description of the dynamics can 
be further improved by introducing diffusion in the Hilbert space  ${\cal S}_Q$.
We consider that the evolutions of $ q^{[a]}_\alpha$ and $ q^{[b]}_\alpha$ now read 
\begin{eqnarray}
d q^{[a]}_\alpha &=& d q^a_\alpha + d \xi^{[2]}_\alpha , \nonumber \\
d {q^{[b]}_\alpha}^* &=& d {q^b_\alpha}^* + d \eta^{[2]}_\alpha , \nonumber 
\end{eqnarray}
where $d \xi^{[2]}_\alpha$ and $d \eta^{[2]}_\alpha$ correspond to two sets 
of stochastic gaussian variables (following Ito rules of stochastic calculus \cite{Gar85}) 
with mean values equal to zero and variances verifying 
\begin{eqnarray}
d\xi^{[2]}_\alpha  d\xi^{[2]}_\beta &=& d \omega_{\alpha \beta} , \label{eq:dxx0}  \\
d\eta^{[2]}_\alpha d\eta^{[2]}_\beta &=& d \sigma_{\alpha \beta} , \label{eq:dxx1}  \\
d\xi^{[2]}_\alpha  d\eta^{[2]}_\beta &=& 0 \label{eq:dxx}
\end{eqnarray}
We assume that $d \omega_{\alpha \beta}$ and $d  \sigma_{\alpha \beta}$ are proportional to 
$dt$. Note that equation (\ref{eq:dxx}) reflects that the stochastic contributions to the evolutions of $\left| Q_a \right>$
and $\left< Q_b \right|$ are independent. The advantage of introducing the Monte-Carlo method 
can be seen in the average 
evolutions of the states. Keeping only linear terms  in $dt$ in eq. (\ref{eq:statevaria}) gives for instance 
\begin{eqnarray}
\overline {\left| dQ_a \right>} &=& \left\{\sum_\alpha d q^{a}_\alpha A_\alpha + \sum_{\alpha < \beta} 
d \omega_{\alpha \beta } \left( A_\alpha A_\beta + A_\beta A_\alpha \right) \right\}\left| Q_a \right>. 
\label{eq:dq1dq2}
\end{eqnarray}
In the previous section, we have recalled that using trial state leads to an approximate treatment 
of the dynamics associated to effective Hamiltonian which can only be written as a linear superposition of the $A_\alpha$.
Last expression emphasizes that, while the states remain in the ${\cal S}_Q$ space during the stochastic process, 
the average evolution can now simulate the evolution with an effective Hamiltonian containing not only
linear but also quadratic in $A_\alpha$. 

The goal is now to take advantage of this generalization and reduce further the distance between the 
average evolution and the exact one. The most natural generalization of the last section is to
minimize the average action 
\begin{eqnarray}
S = \overline{Tr\left( \left\{ i \hbar \partial^\triangleright_t - i \hbar \partial^\triangleleft_t 
- H  \right\} D \right)} ,
\label{eq:actionaver}
\end{eqnarray} 
with respect to the variations of different parameters, i.e. 
$\delta q^{a}_\alpha$, $\delta {q^{b}_\alpha}^*$,  $\delta \omega_{\alpha \beta}$
and $\delta \sigma_{\alpha \beta}$. In the following, 
a formal solution of the minimization procedure is obtained. We show that variational principles applied 
to stochastic process generalize the deterministic case by imposing that not only that 
expectation values $\left< A_\alpha \right>$ but also the second moments 
$\left< A_\alpha A_\beta \right>$, follow the exact Ehrenfest evolution.

\subsubsection{Effective Hamiltonian dynamics deduced from the minimization}

The variations with respect to $\delta {q^{b}_\alpha}^*$ and $\delta \sigma_{\alpha \beta}$ 
give two sets of coupled equations between $dq^a_\alpha$ and $d\omega_{\alpha \beta}$. 
The formal solution of the minimization can however be obtained by making an appropriate change on the variational 
parameters prior to the minimization. In the following, we introduce the notation 
$B_{\nu} = A_\alpha A_\beta + A_\beta A_\alpha$ where $\nu$ denotes $(\alpha,\beta)$ with $\alpha < \beta$. Starting from the general form of the effective evolution
(\ref{eq:dq1dq2}), we  dissociate the part which contributes to the evolution of the $\left<A_\alpha \right>$ 
from the rest. This could be done by introducing the projection operator 
${\cal P}_1$. Equation (\ref{eq:dq1dq2}) then reads
\begin{eqnarray}
\overline {\left| d Q_a \right>} &=& \left\{\sum_\alpha  d {z^a_\alpha} A_\alpha + \sum_{\nu} d \omega_{\nu} 
(1-{\cal P}_1)B_\nu
\right\}\left| Q_a \right>,
\label{eq:dq1dq2new}
\end{eqnarray}  
where the new set of parameters $d {z^a_\alpha}$ is given by 
\begin{eqnarray}
d {z^a_\alpha} = d {q^a_\alpha} + \sum_{\beta \nu}  
d \omega_{\nu} M^{-1}_{\alpha \beta} \left< Q_b \left| A_\beta B_\nu  \right| Q_a \right>.
\end{eqnarray}  
Similarly, the average evolution $\left< dQ_b \right|$ transforms into
\begin{eqnarray}
\overline {\left< d Q_b \right|} &=& \left< Q_b \right| \left\{\sum_\alpha  d {z^b_\alpha}^* A_\alpha 
+ \sum_{\nu} d \sigma_{\nu} B_\nu(1-{\cal P}_1)
\right\},
\label{eq:dq1dq2new2}
\end{eqnarray}  
where $d {z^b_\alpha}$ is given by 
\begin{eqnarray}
d {z^b_\alpha}^* = d {{q^b_\alpha}^*} + \sum_{\beta \nu}  d \sigma_{\nu} 
 \left< Q_b \left|  B_\nu A_\beta \right| Q_a \right>M^{-1}_{\beta \alpha}.
\end{eqnarray}
In the following, we write $B'_\nu = (1-{\cal P}_1)B_\nu$ and $B''_\nu = B_\nu(1-{\cal P}_1)$. 
The great interest of this transformation is to have $\left< A_\alpha B'_\nu \right> = 0$ and 
$\left< B''_\nu A_\alpha \right>=0$ for all $\alpha$
and $\nu$. Accordingly, the variations with 
respect to $\delta {{z^b_\alpha}^*}$ and $\delta {{z^a_\alpha}}$ lead to 
\begin{eqnarray}
\left\{
\begin{array} {c}
i \hbar \overline{\left< Q_b \left| A_\alpha \right| d Q_a \right>} = \left< Q_b \left| A_\alpha  H \right| Q_a \right> \\
\\
i \hbar \overline{\left< d Q_b \left| A_\alpha \right| Q_a \right>} = \left< Q_b \left|  H A_\alpha \right| Q_a \right>,
\end{array}
\right.
\end{eqnarray}
which gives closed equations for the variations $dz^a_\alpha$ and ${dz^b_\alpha}^*$
which are decoupled from the evolution of $d \omega_{\nu} $ and $d \sigma_{\nu}$. These equations
are identical to the ones derived in step 1 and can be again inverted as 
\begin{eqnarray}
\sum_\alpha  d {z^a_\alpha} A_\alpha \left| Q_a \right> = \frac{dt}{i\hbar} {\cal P}_1 H \left| Q_a \right>, \\
\left< Q_b \right| \sum_\alpha  d {z^b_\alpha}^* A_\alpha  = -\frac{dt}{i\hbar} \left< Q_b \right|
H {\cal P}_1. 
\end{eqnarray}  
On the other hand, the variations with respect to $\delta \sigma_\nu$ and $\delta \omega_\nu$ lead to
\begin{eqnarray}
\left\{
\begin{array} {c}
i \hbar \overline{\left< Q_b \left| B''_\nu \right| d Q_a \right>} = \left< Q_b \left| B''_\nu  H \right| Q_a \right>, \\
\\
i \hbar \overline{\left< d Q_b \left| B'_\nu \right| Q_a \right> } = \left< Q_b \left|  H B'_\nu \right| Q_a \right>,
\end{array}
\right.
\end{eqnarray} 
which again gives closed equations for $d \omega_\nu$ and $d \sigma_\nu$.
These equations can be formally solved by introducing the two projectors 
${\cal P}_2$ and ${\cal P}'_2$ associated respectively to the subspaces of operators $B_\nu(1-{\cal P}_1)$
and $(1-{\cal P}_1)B_\nu$. ${\cal P}_2$ differs from ${\cal P}'_2$ due to the fact that $B_\nu$ operators 
and $A_\alpha$ operators do not a priori commute.
Then, the effective evolution given by eq. (\ref{eq:dq1dq2}) becomes
\begin{eqnarray}
\overline{\left| dQ_a \right>} &=& \frac{dt}{i\hbar} 
\left(\sum_\alpha  d {z^a_\alpha} A_\alpha + (1-{\cal P}_1) \sum_{\nu} d \omega_{\nu} B_\nu \right) \left| Q_a \right> \nonumber \\
&=&  \frac{dt}{i\hbar} \left( {\cal P}_1 + {\cal P}_2 \right)H \left| Q_a \right>, 
\end{eqnarray}   
while 
\begin{eqnarray}
\overline{\left< dQ_b \right|} &=& -\frac{dt}{i\hbar} \left< Q_b \right| H\left( {\cal P}_1 + {\cal P}'_2 \right). 
\end{eqnarray}
In both cases, the first part corresponds to the projection of the exact dynamics on the space of 
observable $\left< A_\alpha\right>$.
The second term corresponds to the projection on the subspace of the observable $\left< A_\alpha A_\beta \right>$ 
"orthogonal" to the space of the $\left< A_\alpha \right>$.

\subsubsection{Interpretation in terms of observable evolution}

The variation with respect to an enlarged set of parameters does a priori completely determine the deterministic 
and stochastic evolution of the two trial state vectors. The associated average Schroedinger evolution 
corresponds to a projected dynamics. 
The interpretation of the solution obtained by variational principle 
is rather clear in terms of observable evolution. Indeed, from the two 
variational conditions, we can easily deduce that 
\begin{eqnarray}
\overline{ d\left< A_\alpha  \right>} &=& \left< \left[ A_{\alpha}, H \right] \right>, \nonumber \\
\overline{d\left< B_\nu  \right>} &=&  \frac{dt}{i\hbar} \left< \left[ B_{\nu}, H \right] \right>. \nonumber 
\end{eqnarray}
In summary, using the additional parameters associated with the stochastic contribution as variational parameters
for the average action given by eq. (\ref{eq:actionaver}), one can further reduce the distance between the 
simulated evolution and the exact solution. When gaussian noises are used, this 
is equivalent to impose that the evolution of the correlations between operators $A_\alpha$ obtained by averaging over 
different stochastic trajectories also matches the exact evolution.

\subsection{Step 3: Generalization}

If the Hamiltonian $H$ applied to the trial state can be written as a quadratic Hamiltonian in terms of $A_\alpha$ and 
if the trial states form an overcomplete basis of the total Hilbert space, then the above procedure 
can provide with an exact stochastic reformulation of the problem.
If it is not the case, the above methods can be generalized  by introducing higher 
order stochastic variables. Considering now the more general form
\begin{eqnarray}
\left\{
\begin{array} {c}
\delta q^{[a]}_\alpha = \delta q^a_\alpha + \delta \xi^{[2]}_\alpha + \delta \xi^{[3]}_\alpha+ \cdots \nonumber \\  
\delta {q^{[b]}_\alpha}^* = \delta {q^b_\alpha}^* + \delta \eta^{[2]}_\alpha + \delta \eta^{[3]}_\alpha+ \cdots \nonumber   
\end{array}
\right.
\end{eqnarray}  
we suppose now that the only non vanishing moments for $d\xi^{[k]}_\alpha$ and $d\eta^{[k]}_\alpha$ are the moments 
of order $k$ (which are then assumed to be proportional to $dt$). 
For instance, we assume that $d\xi^{[3]}_\alpha$ verifies 
\begin{eqnarray}
\overline{d\xi^{[3]}_\alpha} = \overline{d\xi^{[3]}_\alpha d\xi^{[3]}_\beta}&=& 0 , \\
\overline{d\xi^{[3]}_\alpha d\xi^{[3]}_\beta d\xi^{[3]}_\gamma} &\neq& 0.
\end{eqnarray}
Then without going into details, we can easily generalize the method presented in step 2 and deduce that
the average evolutions of the trial states will be given by
\begin{widetext}
\begin{eqnarray}
\overline{\left| d Q_a \right>}  
&=& \frac{dt}{i\hbar}\left\{ {\cal P}_1  + {\cal P}_2 + {\cal P}_3 + \cdots \right\} H \left| Q_a \right> \nonumber \\
\overline{\left< d Q_b \right|}  &=& -\frac{dt}{i\hbar}\left< Q_b \right|  H \left\{ {\cal P}_1  + {\cal P}'_2 + {\cal P}'_3 + \cdots \right\} \nonumber
\label{eq:stocgen}
\end{eqnarray}
where the first terms contain all the information on the evolution of the $\left< A_\alpha \right>$, the 
second terms contain all the information on the evolution of the $\left< A_\alpha A_\beta \right>$ which is not
accounted for 
by the first term, the third terms contain all the information on the evolution of the $\left< A_\alpha A_\beta A_\gamma 
\right>$ which is not contained in the first two terms, ... 
The procedure described here gives an exact Monte-Carlo formulation of a given problem if the Hamiltonian 
$H$ applied on $\left| Q_a \right>$ or $\left< Q_b \right|$ 
can be written as a polynomial of $A_\alpha$. If the polynomial is of order $k$, then the sum stops 
at ${\cal P}_k$.
\end{widetext}

\subsection{Summary and discussion on practical applications}

In this section, we propose a systematic method to replace a general quantum problem by stochastic processes within 
a restricted class of trial state vectors associated to a set of observable $A_\alpha$. Using variational techniques, 
we show that a hierarchy of stochastic approximations can be obtained. This method insures that at
the level $k$ of the hierarchy, all moments of order $k$ or below of the observable $A_\alpha$ evolve 
according to the exact Ehrenfest equation over a single time step. The Monte-Carlo formulation might  
becomes exact if the Hamiltonian applied to the trial state writes as a 
polynomial of the $A_\alpha$ operators. 
  
Aside of the use of variational techniques, we end up with the following important conclusion:
{\it Given an initial density $D=\left| Q_a \right>\left< Q_b \right|$ where both states belongs to a
given class of trial states associated to a set of operators $A_\alpha$, we can always 
find a Monte-Carlo process which preserves the specific form of $D$ and insures that expectations values of 
all moments of the $A_\alpha$ up to a certain order $k$ evolve in average according to the Ehrenfest theorem 
associated to the exact Hamiltonian at each time step and along each trajectory.}  

This statement will be referred to as the "existence theorem" in the following. Such a general statement 
is very useful
in practice to obtain stochastic processes.
Indeed, the use of variational techniques might become rather complicated 
due to the large number of degrees of freedom involved. 
An alternative method is to take advantage of the natural link made
between the average effective evolution deduced from the stochastic evolution 
and the phase-space dynamics. Indeed, according 
to the existence theorem, we know that at a given  
level $k$ of approximation, the dynamics of each trial state can be simulated by an average effective 
Hamiltonian insuring that all moments of order $k$ or below matches the exact evolution.
In general, it is easier to express the exact evolution of the moments and then 'guess' the 
associated SSE.

The second important remark is that the above method can also be used to provide a SSE which maintains 
$Tr(D)=1$ along each path. In that case, the action (\ref{eq:actionaver}) can still be used 
but the density $D$ should be replaced by  
\begin{eqnarray}
D = \frac{\left| Q_a \right>\left< Q_b \right|}{\left< Q_b \left.  \right| Q_a \right>}.
\end{eqnarray}     
All the equations given above can then be equivalently derived. The main difference being that 
$A_\alpha$ is now replaced by  $(A_\alpha - \left< A_\alpha \right>)$ in the variations of the states 
(equation (\ref{eq:statevaria})). As we will see, the possibility to use constant trace 
formulation usually simplifies the use of the Ehrenfest theorem to guess the involved stochastic equations.

\section{Applications}

In the following, we give few examples which have been studied recently in the context of Monte-Carlo techniques 
to illustrate how formal results obtained in previous section can be applied.  
In each case, the link between the phase-space evolution and the 
associated SSE is pointed out.

\subsection{The free wave-packet expansion}

We first consider a system initially in the ground state of a harmonic oscillator described by the 
Hamiltonian $H_0 = \hbar \omega  \left( a^+ a + \frac{1}{2} \right)$ 
where $a$ and $a^+$ are the usual creation/annihilation operators with $[a,a^+]=1$.
Denoting by $\eta = m\omega/\hbar$, the initial wave-packet corresponds to a gaussian wave-function 
of width $<x^2> =\frac{1}{2\eta}$.   
At time $t=0$, the harmonic constraint is released. During the free expansion, 
the wave-packet remains gaussian and verifies \cite{Coh77} 
\begin{eqnarray}
<x^2> =\frac{1}{2\eta} +  \frac{\hbar^2 \eta}{2m^2} t^2,
\label{eq:exactdif}
\end{eqnarray}
where $m$ is the mass of the system.

\subsubsection{Stochastic formulation}

We show that this free expansion can be simulated by quantum diffusion 
between gaussian wave-functions of fixed widths. 
Let us assume that, at a given 
intermediate time step of the stochastic process, the density is given by averaging over
densities of the form (which includes the initial condition) 
\begin{eqnarray}
D = \frac{\left| \alpha  \right> \left< \beta  \right|}{\left< \beta \left.  \right| \alpha \right>},
\label{eq:denscoh} 
\end{eqnarray} 
where $\left| \alpha  \right>$ and $\left< \beta  \right|$ are Bargmann states of the initial oscillator 
 defined as $\left| \alpha \right> = e^{\alpha a^+} \left| 0 \right>$\cite{Gar00}. These states verify 
$a\left| \alpha  \right> = \alpha \left| \alpha  \right>$ and 
$\left< \beta  \right|a^+ = \left< \beta  \right| \beta^*$. We use Bargmann states instead of coherent states 
to simplify the discussion. In particular, we have the transformation properties 
\begin{eqnarray}
\begin{array}{ccc}
\left| \alpha + d\alpha \right> = e^{d\alpha a^+} \left| \alpha \right>,~~\left< \beta + d\beta  \right|  =
\left< \beta \right| e^{d\beta^* a},
\end{array}
\label{eq:stob}
\end{eqnarray} 
which illustrates that the generator of transformations between Bargmann states are the $a$ and $a^+$ operators. 

Once the harmonic constraint is released the Hamiltonian becomes the free Hamiltonian 
\begin{eqnarray}
H = \frac{p^2}{2m} = -\frac{\hbar \omega}{4} (a^+-a)^2.
\end{eqnarray}
The action of $H$ on $\left| \alpha  \right>$ and $\left< \beta \right|$ can be written 
respectively as
\begin{eqnarray}
H\left| \alpha  \right> &=& -\frac{\hbar \omega}{4}  ({a^+}^2 -2 \alpha a^+ + \alpha^2)\left| \alpha  \right>, \nonumber \\
\left< \beta \right|H &=& -\frac{\hbar \omega}{4}  \left< \beta  \right|({a}^2 -2 \beta^* a + {\beta^*}^2), \nonumber
\end{eqnarray}
which corresponds to polynomial of order $2$ in $a^+$ and $a$ respectively. Therefore, 
we expect that the exact dynamics can be simulated with gaussian stochastic process between densities 
given by eq. (\ref{eq:denscoh}). The associated stochastic evolution 
of $\alpha$ and $\beta^*$ are written as 
\begin{eqnarray}
\left\{
\begin{array} {ccc}
d\alpha &=& \overline{d\alpha} + d\xi^{[2]}, \nonumber \\
d\beta^* &=& \overline{d\beta^*} + d\eta^{[2]}, 
\end{array}
\right.
\label{eq:stocab}
\end{eqnarray}   
where $\overline{d\alpha}$ and $\overline{d\beta^*}$ correspond to the deterministic part of the evolution, 
while $d\xi^{[2]}$ and $d\eta^{[2]}$ are two independent gaussian stochastic variables with zero mean 
and following Ito rules \cite{Gar85}. 
In order to precise the nature of the stochastic evolution, one can use the fact that the first 
and second moments of the $a$ and $a^+$ operators follow in average the exact evolution over one time step.
Noting that, for $D$ given by eq. (\ref{eq:denscoh}), we have 
$\left< a^+ \right> = Tr(a^+ D) =\beta^*$ and $\left< a \right> = Tr(a D) =\alpha$, the Ehrenfest theorem applied 
to $\left< a^+ \right>$ and $\left< a \right>$ gives
the two equations
\begin{eqnarray}
\overline{d \alpha}  &=& \overline{d \beta^*} = i  \frac{\hbar\eta}{2m}(\beta^* -\alpha).
\end{eqnarray}
The application of the Ehrenfest theorem respectively to $\left<\right. a^2 \left.\right>$, $\left<\right.  {a^+}^2 \left.\right>$ and 
$\left< a^+ a \right>$ gives
\begin{eqnarray}
\overline{d{(\alpha^2)}}  &=&    \frac{i\hbar\eta}{m}dt \left\{(\beta^* -\alpha)\alpha  + \frac{1}{2} \right\}, \nonumber  \\
\overline{d({\beta^*}^2)} &=&    \frac{i\hbar\eta}{m}dt \left\{(\beta^* -\alpha)\beta^* - \frac{1}{2} \right\}, \nonumber \\
\overline{d {(\alpha \beta^*)}} &=& \frac{i\hbar\eta}{m}dt \left\{(\beta^* -\alpha)(\alpha+\beta^*) \right\}. \nonumber 
\end{eqnarray}
Last condition is automatically fulfilled if we assume that $d\xi^{[2]}$
and $d\eta^{[2]}$ are independent stochastic variables. According to Ito rules, we also have 
\begin{eqnarray}
\overline{d({\alpha^2})} &=&  2 \alpha \overline{d \alpha} +   d\xi^{[2]}d\xi^{[2]} , \nonumber \\
\overline{d({\beta^*}^2)} &=& 2 \beta^* \overline{d \beta^*} + d\eta^{[2]}d\eta^{[2]},  \nonumber
\end{eqnarray}
from which we deduce 
\begin{eqnarray}
d\xi^{[2]}d\xi^{[2]}   &=& -d\eta^{[2]}d\eta^{[2]} = \frac{i\hbar\eta}{2m}dt .
\end{eqnarray}
A stochastic evolution of $\alpha $ and $\beta^*$ compatible 
with the above conditions is then
\begin{eqnarray}
d\alpha = i\hbar \left\{\frac{\eta}{2m}(\beta^* -\alpha) + \sqrt{\frac{\eta}{2m}}dW_1 \right\} , \\
d\beta^*  = i\hbar \left\{\frac{\eta}{2m}(\beta^* -\alpha) + \sqrt{\frac{\eta}{2m}}dW_2 \right\} , 
\label{eq:stocfree}
\end{eqnarray}  
where $dW_1$ and $dW_2$ are independent stochastic variables with zero mean and 
$dW_1 dW_1 = -dW_2dW_2 =\frac{dt}{i\hbar}$. The above $c$-number dynamic 
defines a stochastic evolution of the trial states through the relation (\ref{eq:stob}) 
or equivalently can be interpreted as a diffusion process in the 
space of densities given by expression (\ref{eq:denscoh}).

\subsubsection{Recovering the exact dynamics from the average evolution}

Let us now check that the stochastic dynamics properly describes 
 the exact evolution. For this, it is convenient 
to consider the stochastic complex variables defined as $X=Tr(Dx)$ and $P=Tr(Dp)$. 
For densities given by  eq. (\ref{eq:denscoh}), $X$ and $P$ are related to $\alpha $
and $\beta^*$ through the relations
\begin{eqnarray}
\left\{
\begin{array} {ccc}
X &=& ~~\frac{1}{\sqrt{2\eta}}(\alpha + \beta^*) , \nonumber \\
\\
P &=& i\hbar \sqrt{\frac{\eta}{2}} (\beta^* - \alpha). \nonumber
\end{array}
\right.
\end{eqnarray} 
Accordingly, $X$ and $P$ are stochastic complex variables. 
Starting from eq. (\ref{eq:stocfree}), 
we obtain
\begin{eqnarray}
\left\{
\begin{array} {ccc}
d X &=& \frac{P}{m}dt +  d \chi _1,  \\
\\
d P &=& d\chi _2,
\end{array}
\right.
\label{eq:dynpq}
\end{eqnarray}
where the new stochastic variables $d \chi _1$ and $d \chi _2$ verify 
\begin{eqnarray}
\overline{d\chi _1} &=& \overline{d\chi _2} = 0, \nonumber \\
\overline{d\chi _1d\chi _1} &=& \overline{d\chi _2 d\chi _2} = 0, \nonumber \\
\overline{d\chi _1 d \chi _2} &=& \frac{\hbar^2 \eta}{2m}dt .   \nonumber
\end{eqnarray}
The Langevin equations (\ref{eq:dynpq}) identifies with the classical equation of 
motion of the free particle with extra complex gaussian noises. 
Assuming that initially $X(0)=P(0)=0$,
and using Ito rules of stochastic calculus, we deduce that
\begin{eqnarray}
\overline{X^2(t)} = \frac{\hbar^2 \eta}{2m^2}t^2. \nonumber
\end{eqnarray}
Since along each stochastic path $Tr(Dx^2) = \frac{1}{2\eta} + X^2$, we finally end up with
\begin{eqnarray}
\overline{Tr(Dx^2)} = \frac{1}{2\eta}  + \frac{\hbar^2 \eta}{2m^2}t^2, \nonumber
\end{eqnarray}
which is nothing but the exact result given by eq. (\ref{eq:exactdif}). 

With this simple example, we have illustrated how a quantum problem can be transformed 
into a Monte-Carlo process between densities formed of two Bargmann states. A remarkable 
aspect of the method used here is that at no time it is needed to minimize directly the 
action. We only took advantage of both the existence theorem and Ehrenfest evolution 
to obtain the Stochastic Schroedinger evolution of trial states. It should however 
be kept in mind that the above stochastic process is intimately connected to 
the projection technique described in previous section.

\subsection{The Kerr oscillator}
\label{sect:kerr}

As a second example, we consider the model Hamiltonian known as the Kerr oscillator
\begin{eqnarray}
H= \hbar \omega a^+ a +\frac{1}{2} \hbar K (a^+)^2 a^2.
\end{eqnarray}
Using 
either the positive-P representation \cite{Gar00} or more recent works based 
on quantum-field theoretical techniques \cite{Pli03}, it can be shown that the quantum 
problem of a system evolving with $H$ can be replaced by an ensemble of stochastic c-number 
evolutions. 
Let us follow the same strategy as in previous example, and assume that at a given time step
the density can be recovered by averaging over densities given by eq. (\ref{eq:denscoh}). Using
similar arguments as before, we know that the exact Hamiltonian dynamics can be simulated 
by a gaussian diffusion process on $\alpha$ and $\beta^*$, which have the general form 
given by eq. (\ref{eq:stocab}). The Ehrenfest theorem applied on $\left< a \right>$ and $\left< a^+ \right>$
gives directly 
\begin{eqnarray}
\overline{d \alpha}  &=&  -i dt \left\{\omega +  K \left[ \beta^*\alpha\right] \right\}  \alpha, \nonumber  \\
\overline{d \beta^*}  &=&  ~~i dt \left\{\omega +  K \left[ \beta^*\alpha\right] \right\}  \beta^*, \nonumber 
\end{eqnarray}
while the evolutions of $\left<\right. a^2 \left.\right>$, $\left<\right.  {a^+}^2 \left.\right>$ and 
$\left< a^+ a \right>$ lead respectively to
\begin{eqnarray}
\overline{d{\alpha^2}}  &=&   -2i dt \left\{\omega +  K \left[ \beta^*\alpha\right] \right\}  \alpha^2 -iKdt \alpha^2 , 
\nonumber  \\
\overline{d{\beta^*}^2} &=&  ~~2i dt \left\{\omega +  K \left[ \beta^*\alpha\right] \right\}  {\beta^*}^2 + 
iKdt \beta^2 , \nonumber  \\
\overline{d {(\alpha \beta^*)}} &=& ~~0 .  \nonumber 
\end{eqnarray}
From the two sets of equations, we deduce 
\begin{eqnarray}
d\xi^{[2]}d\xi^{[2]}   &=&  -iKdt  \alpha^2 , \nonumber \\
d\eta^{[2]}d\eta^{[2]} &=&   ~~iKdt {\beta^*}^2. \nonumber 
\end{eqnarray}
Therefore, a stochastic Monte-Carlo process compatible with the above condition is given by   
\begin{eqnarray}
d \alpha  &=&  -idt \left\{\omega +  \alpha K \left[ \beta^*\alpha\right] \right\} 
+ \sqrt{K \hbar} \alpha ~dW_1  \nonumber \\
d \beta^* &=&   ~idt \left\{\omega +  K \left[ \beta^*\alpha\right] \right\} 
 \beta^* + \beta^* \sqrt{K \hbar}  dW_2  \nonumber 
\end{eqnarray}
where again $dW_1$ and $dW_2$ are independent stochastic variables with zero mean and 
$dW_1 dW_1 =-dW_2dW_2 =\frac{dt}{i\hbar}$. The above stochastic process is exactly the same as the one derived in 
ref.  \cite{Pli03} using a completely different technique. Indeed, 
although the presented framework uses phase-space
arguments, it has a strong connection with the Hamiltonian dynamics.
A great advantage here is that the guess of the Monte-Carlo process is rather straightforward
compared to ref. \cite{Pli03}. Note that above stochastic equations are non-linear stochastic differential equation 
which might be numerically difficult to handle \cite{Pli01}.     

\subsection{The generalized Kerr Hamiltonian}

Up to now, we have presented examples where the exact Hamiltonian can be recovered using gaussian noises.
To illustrate further the simplicity of the method, we consider the following Hamiltonian  
\begin{eqnarray}
H= \hbar \omega a^+ a +\frac{1}{2} \hbar K_1 (a^+)^2 a^2 + \frac{1}{6} \hbar K_2 (a^+)^3 a^3,
\label{eq:kerr3}
\end{eqnarray}
which will be referred hereafter as the generalized Kerr Hamiltonian.
According to the existence theorem, in order to simulate exactly the Hamiltonian, 
it is necessary to include an extra contribution to the c-number evolution
\begin{eqnarray}
d\alpha &=& \overline{d\alpha} + d\xi^{[2]} + d\xi^{[3]}, \nonumber \\
d\beta^* &=& \overline{d\beta^*} + d\eta^{[2]} + d\eta^{[3]},  \nonumber
\end{eqnarray}    
where first and second moments of $d\xi^{[3]}$ and $d\eta^{[3]}$ cancel out while third moments 
are proportional to $dt$. As in previous section, the use of the Ehrenfest theorem for the
first and second moments leads respectively to 
\begin{eqnarray}
\overline{d \alpha} &=&  - idt \left\{\omega +  K_1 \left[ \beta^*\alpha\right] + \frac{1}{2} K_2 
\left[ \beta^*\alpha\right]^2 \right\}  \alpha , \nonumber \\
\overline{d \beta^*} &=& ~~idt   \left\{\omega +  K_1 \left[ \beta^*\alpha\right] + \frac{1}{2} K_2 \left[ \beta^*\alpha\right]^2 
\right\}  \beta^* , \nonumber
\end{eqnarray} 
and 
\begin{eqnarray}
d\alpha^{[2]} d\alpha^{[2]}  &=& -i dt {\alpha}^2(K_1  + K_2 \left[ \beta^*\alpha\right]) , \nonumber \\
d{\beta^*}^{[2]} {d\beta^*}^{[2]} &=& i dt {\beta^*}^2(K_1  + K_2 \left[ \beta^*\alpha\right]). \nonumber 
\end{eqnarray}
In order to determine the extra contribution, one should use the third moments. The evolution of $\left<\right. 
{a^+}^3 \left. \right>$ 
gives for instance
\begin{widetext}
\begin{eqnarray}
 \overline{d (\beta^*)^3} =  i {\beta^*}^3 \left\{
3 \left[ \omega +  K_1 \left[ \beta^*\alpha\right] + \frac{1}{2} K_2 \left[ \beta^*\alpha\right]^2 \right]
+ 3  \left(  K_1 + K_2 \left[ \beta^*\alpha\right]\right)+ K_2 \right\}. \nonumber 
\end{eqnarray}
Using the identity 
\begin{eqnarray}
\overline{d (\beta^*)^3}  = 3 {\beta^*}^2 \overline{d \beta^*} + 3 \beta^* d{\beta^*}^{[2]} {d\beta^*}^{[2]} + 
d{\beta^*}^{[3]} {d\beta^*}^{[3]}{d\beta^*}^{[3]} , \nonumber
\end{eqnarray}
\end{widetext}
gives   
\begin{eqnarray}
d{\beta^*}^{[3]}d{\beta^*}^{[3]}d{\beta^*}^{[3]} &=& idt \hbar K_2 {\beta^*}^3. \nonumber
\end{eqnarray}
Similarly, the evolution of $\left< \right. {a}^3 \left. \right>$ gives
\begin{eqnarray}
d{\alpha}^{[3]}d{\alpha}^{[3]}d{\alpha}^{[3]} &=& -idt \hbar K_2 {\alpha}^3. \nonumber
\end{eqnarray} 
Therefore, the exact evolution of the generalized Kerr oscillator can be simulated with coherent states 
using the stochastic process on the c-numbers
\begin{widetext}
\begin{eqnarray}
d \alpha  =  -i dt \left\{\omega +  
K_1 \left[ \beta^*\alpha\right] + \frac{1}{2} K_2 \left[ \beta^*\alpha\right]^2 \right\}  \alpha + 
\left( d\xi_2 + d\xi_3 \right) \alpha ,
 \nonumber \\
d \beta^*  =   i dt\left\{\omega +  K_1 \left[ \beta^*\alpha\right] + 
\frac{1}{2} K_2 \left[ \beta^*\alpha\right]^2  \right\}  \beta^* + \left( d\eta_2 + d\eta_3 \right) \beta^* ,
\nonumber
\end{eqnarray}
\end{widetext}
where the $d\xi_i$ and $d\eta_i$ are independent stochastic variables verifying 
\begin{eqnarray}
\overline{d\xi_i} &=& \overline{d\eta_i} =0, \nonumber \\
\overline{d\xi_3 d\xi_3} &=& \overline{d\eta_3 d\eta_3}  = 0, \nonumber \\
\overline{d\xi_2 d\xi_2} &=& -\overline{d\eta_2 d\eta_2}  = -i dt (K_1  + K_2 \left[ \beta^*\alpha\right]) , \nonumber \\
\overline{d\xi_3 d\xi_3 d\xi_3} &=& -\overline{d\eta_3 d\eta_3 d\eta_3} = -idt K_2. \nonumber 
\end{eqnarray} 
The exactness of the above stochastic process can be checked using the fact that for any density given by equation 
(\ref{eq:denscoh}) and the stochastic evolution of $\alpha$ and $\beta^*$ given above, we have the average relation
\begin{eqnarray}
\overline{e^{d\alpha (a^+-\left< a^+ \right>)} D + D e^{d\beta^* (a-\left< a \right>)}} = 
D + \frac{dt}{i\hbar} \left[H,D\right] , 
\end{eqnarray}
where $H$ is the total Hamiltonian given by eq. (\ref{eq:kerr3}). This relation shows 
that the average evolution identifies with the exact evolution for a given $D$ and a single time step.  
Note that here the use of $(a^+-\left< a^+ \right>)$ and $(a-\left< a \right>)$ 
is related to the constant trace, 
$Tr(D)=1$, imposed along each paths.   
 
\subsection{Interacting Boson systems}
\label{sect:boson}

In recent years, several works have demonstrated that bosons \cite{Car01} or fermions \cite{Jul02}
interacting through a two-body potential can be simulated exactly by a Monte-Carlo method using 
simple trial state vector. Let us first consider 
the general case of N interacting bosons described by the general two-body Hamiltonian 
\begin{eqnarray}
H = \sum_{ij} \left< i \left| T \right| j \right> a^+_i a_j + 
\frac{1}{2} \sum_{ijkl} \left< ij \left| v_{12} \right| lk \right> a^+_i a^+_j a_l a_k ,
\end{eqnarray}  
where the $a_i$ and $a^+_i$ correspond to a complete set of creation/annihilation operators which 
obey bosons commutation rules. 
It has been shown in ref. \cite{Car01}, that Bose-Einstein condensates wave-functions
can be used as trial wave-functions. In that case, the generator of the transformations 
between two non-orthogonal wave-function are the single-particle operators $a^+_ia_j$. 
Since $H$ can be written as a quadratic polynomial in terms of these operators, we can have guessed 
the existence of such a  stochastic reformulation.     
We show here that the proposed method  also leads  
to an exact reformulation of the exact problem in terms of jumps between densities   
\begin{eqnarray}
D &=& \left| N: \phi_a \right> \left< N: \phi_b \right| ,
\end{eqnarray} 
where both $\left| N: \phi_a \right>$ and $\left| N: \phi_b \right>$ stand for two Bose-Einstein condensates wave-function
and where
$Tr(D)=1$ along each path.

\subsubsection{Phase-space stochastic dynamics}

Let us assume that the system is at initial time $t_0$ in a specific Bose condensate wave-function 
$\left| N: \phi \right>$, where $\left| \phi \right>$ is the associated single-particle state.
Due to the nature of the initial state, all the information is contained in the one-body 
degrees of freedom. It is then convenient to define the one-body density 
matrix $\left< j \left| \rho_1 \right| i \right> = \left< a^+_i a_j \right>$, whose matrix elements 
can be written as
\begin{eqnarray}
\rho_1     &=& N \left| \phi \right> \left< \phi \right| = N P_\phi,
\label{eq:bos1}
\end{eqnarray}
$P_\phi$ being the projector on the single particle state $\left| \phi \right>$.
According to the existence theorem, we know that an exact Monte-Carlo reformulation can be obtained 
using the first and second moments of a complete set of one-body operators. 
This is equivalent to express
the evolution of the one and two-body densities over one time step. Indeed, the two-body density denoted by 
$\rho_{12}$ is defined as
usual as $\left< ij \left| \rho_{12} \right| kl \right>= \left< a^+_k a^+_l a_j a_i \right>$. 
For a pure Bose-Einstein condensate, $\rho_{12}$ reads 
\begin{eqnarray}
\rho_{12} = N(N-1) P_\phi(1) P_\phi(2),
\label{eq:bos2}
\end{eqnarray}        
where $P_\phi(i)$ means that the projection is made on the $i^{th}$ particle. The evolution of $\rho_1$ and $\rho_{12}$
over one time step are given by the first two  equations of the Bogolyubov-Born-Green-Kirwood-Yvon
(BBGKY) hierarchy \cite{Kir46,Bog46,Bor46}, which reduces in the present case to the two equations of motion
\begin{eqnarray}
i \hbar \frac{d}{dt} \rho_1 &=& \left[ h_{MF}(1), \rho_1 \right] ,  \\
i \hbar \frac{d}{dt} \rho_{12} &=& [h_{MF}(1)+ h_{MF}(2), \rho_{12}] + B_{12}.
\end{eqnarray} 
$h_{MF}$ denotes the mean-field Hamiltonian given by  
\begin{eqnarray}
h_{MF} = T+(N-1) Tr_2(v_{12} P_\phi (2)),
\end{eqnarray}
where $Tr_2$ corresponds to the partial trace on the second particle.
$B_{12}$ is the extra contribution which treats correlations beyond mean-field 
and reads 
\begin{eqnarray}
B_{12} &=& (1-P_\phi(1)) (1-P_\phi(2)) v_{12} \rho_{12} \nonumber \\
&&-  \rho_{12} v_{12} ( 1-P_\phi(1)) (1-P_\phi(2)). 
\end{eqnarray}
Due to this extra contribution, the exact many-body density differs from a pure Bose-Einstein 
wave-function after one time step evolution. 

Starting from the initial densities given by eq. (\ref{eq:bos1}) and (\ref{eq:bos2}),  it is
possible to guess a phase-space stochastic dynamics that simulates exactly the evolution 
of $\rho_1$ and $\rho_{12}$. This is indeed the case if we assume that $P_\phi$ evolves 
according to a stochastic equation that verifies
\begin{eqnarray}
\overline{d P_\phi} &=& \frac{dt}{i\hbar} \left[h_{MF},P_\phi\right], \nonumber \\
\overline{d P_\phi(1) dP_\phi(2)} &=& \frac{dt}{i\hbar} (1-P_\phi(1)) (1-P_\phi(2)) v_{12} P_\phi(1)P_\phi(2)  \nonumber \\
&&-  P_\phi(1) P_\phi(2)  v_{12} ( 1-P_\phi(1)) (1-P_\phi(2)). \nonumber
\end{eqnarray}  
Following ref. \cite{Car01,Jul02} we can write $v_{12}$ as a sum of squares of hermitian one-body operators:  
$v_{12} = \sum_k O_\lambda(1) O_\lambda(2)$. We see that a proper choice for the stochastic evolution of $P_\phi$ is
\begin{widetext}
\begin{eqnarray}
dP_\Phi &=& \frac{dt}{i\hbar} \left[h_{MF},P_\phi  \right] + \sum_{\lambda} d\xi^{[2]}_\lambda (1-P_\phi) O_\lambda P_\phi + \sum_{\lambda} d\eta^{[2]}_\lambda  
P_\phi O_\lambda (1-P_\phi) ,
\label{eq:dphibos}
\end{eqnarray}
\end{widetext}
where $d\xi^{[2]}_\lambda$ and $d\eta^{[2]}_\lambda$ are two sets of independent stochastic variables with mean zero and 
verifying $d\xi^{[2]}_\lambda d\xi^{[2]}_{\lambda'} = \delta_{\lambda\lambda'} \frac{dt}{i\hbar}$  and 
$d\eta^{[2]}_\lambda d\eta^{[2]}_{\lambda'} = -\delta_{\lambda\lambda'} \frac{dt}{i\hbar}$
\subsubsection{Equivalent stochastic process in Hilbert space}
The above phase-space stochastic dynamics can be simulated through a stochastic Schroedinger equation on single-particle 
wave-functions. Then, $P_\phi(t) = \left| \phi \right> \left< \phi \right|$ transforms into a set of couples of 
wave-packets $P_\phi(t+dt) = \left| \phi + d\phi_a \right> \left< \phi +d\phi_b \right|= \left| \phi_a  \right>
\left< \phi_b \right|$,  where states evolves over one time step 
according to   
\begin{eqnarray}
\left\{
\begin{array} {ccc}
d \left| \phi_a \right> &=& \left(\frac{dt}{i\hbar} h_{MF} + \sum_{\lambda} d\xi^{[2]}_\lambda (1-P_\phi) 
O_\lambda \right)\left| \phi_a \right> \nonumber \\
d \left< \phi_b \right| &=& \left< \phi_b \right| \left(-\frac{dt}{i\hbar}h_{MF} + \sum_{\lambda} d\eta^{[2]}_\lambda
O_\lambda (1-P_\phi) \right). \nonumber 
\end{array}
\right.
\end{eqnarray}
where $\left| \phi_a(t_0) \right>=\left| \phi_b(t_0) \right>=\left| \phi \right>$.
It turns out that the above stochastic process preserves $P^2_\phi = P_\phi$ for each trajectory, i.e. 
$\left< \phi_a \left.  \right| \phi_b\right> = 1$. 

This stochastic evolution corresponds to quantum jumps in many-body space which transform the 
initial pure Bose-Einstein condensate into an ensemble of densities given by 
\begin{eqnarray}
D = \left| N: \phi_a \right> \left< N:\phi_b \right|,
\end{eqnarray}  
with $\left< \phi_a \left.  \right| \phi_b\right> = 1$ (which implies $Tr(D)=1$) and is a priori valid only for the first time step.
Let us now assume that we start from one of this many-body densities. It could be shown without difficulty 
that all the above equation remains valid except that now $P_\phi = \left| \phi_a \right> \left< \phi_b \right|$.
Therefore, the single-particle SSE given above can be used to describe the 
long time evolution of the system. Note finally that the Monte-Carlo equations presented here 
differs from the "constant trace"
scheme described in ref. \cite{Car01} in particular due to the use of a common mean-field Hamiltonian in the evolution
of $\left| \phi_a \right>$ and $\left< \phi_b \right|$. 

\subsection{Interacting Fermions systems}
We consider now a system 
of fermions interacting through a two-body Hamiltonian
\begin{equation}
H=\sum_{ij}T_{ij}a_{i}^{+}a_{j}+\frac{1}{4}%
\sum_{ijkl}\left\langle ij\left| v_{12}\right|kl\right\rangle a_{i}^{+}a_{j}^{+}a_{l}a_{k} , 
\label{eq:hamil}
\end{equation}
where the creation/annihilation operators now obey fermionic commutation rules and 
where ${v}_{12}$ is antisymmetric.
Since the derivation of a Monte-Carlo theory is very similar to the boson case, we only give here 
the important steps.

Let us assume that at a given time step, the exact density can be recovered by averaging over an ensemble 
of densities  
\begin{eqnarray}
D = \left| \Phi_a \right>\left< \Phi_b \right|,
\label{eq:phiaphib}
\end{eqnarray}
where both states correspond to Hartree-Fock states. If we denote by $\left\{ \left| \beta_i \right> \right\}_{i=1,N}$ 
and $\left\{ \left| \alpha_i \right> \right\}_{i=1,N}$ the set of $N$ single-particle states, we assume in 
addition that for each couples of Hartree-Fock states, associated singles-particle wave-functions verify  
$\left< \beta_j \left.  \right| \alpha_i \right> = \delta_{ij}$. Accordingly, the one-body density matrix
associated to a given $D$ reads \cite{Low55,Lac05,Lac06-2} 
\begin{eqnarray}
\rho_1 = \sum_i \left| \alpha_i \right> \left< \beta_i \right|.
\label{eq:rhofermion}
\end{eqnarray}     
We can easily verify that $\rho^2_1 = \rho_1$ and $Tr(D)=1$.
For each $D$ given by eq. (\ref{eq:phiaphib}), the two-body density 
is given by $\rho_{12} = (1-P_{12}) \rho_1 \rho_2$ where $P_{12}$ corresponds to the permutation operator. 
The evolution of $\rho_1$ and $\rho_{12}$ over one time step are given by the 
two first equations of the BBGKY hierarchy which reads in that case
\begin{widetext}
\begin{eqnarray}
i \hbar \frac{d}{dt}\rho_1 &=& \left[h_{MF},\rho_1 \right] ,
\label{eq:ro1fermion}\\  
i\hbar \frac{ d }{dt } \rho_{12} &=& \left[ h_{MF}(1)+h_{MF}(2),\rho_{12} \right] \nonumber \\
&+& ( 1 - \rho_1)(1- \rho_2)v_{12} \rho_{1}\rho_{2}
-\rho_{1}\rho_{2}v_{12} ( 1 - \rho_1)(1 - \rho_2),
\label{eq:ro12fermion}
\end{eqnarray}
\end{widetext}
where the mean-field Hamiltonian now reads $h_{MF}= T + Tr_2 (v_{12}\rho_2)$.
This expression identifies with the one obtained for bosons 
except that $P_\phi$ is now replaced 
by the one-body density. 
Again, introducing a complete set of hermitian operators $O_\lambda$, the previous expression can be 
simulated by a stochastic dynamics in phase-space given by eq. (\ref{eq:dphibos}) 
(where again $P_\phi$ is replaced by $\rho_1$) equivalent to a stochastic process for single-particle 
wave-functions given by
\begin{eqnarray}
d \left| \alpha_i \right> &=& \left( \frac{dt}{i\hbar} h_{MF} + \sum_\lambda d\xi^{[2]}_\lambda (1-\rho_1) 
O_\lambda  \right) \left| \alpha_i \right>, \nonumber \\
d \left< \beta_i \right| &=& \left< \beta_i \right| 
\left( -\frac{dt}{i\hbar} h_{MF} + \sum_\lambda d\eta^{[2]}_\lambda  O_\lambda (1-\rho_1) \right). \nonumber 
\end{eqnarray}  
This stochastic equation on single-particle wave-functions, which has been found using a 
different technique in ref. \cite{Jul06}, preserves the property 
$\left< \beta_j \left.  \right| \alpha_i \right> = \delta_{ij}$.
Therefore, it corresponds in many-body space to a Monte-Carlo procedure which 
transforms the initial set of 
densities into another set of densities with identical properties. 

\subsubsection{Possible Extensions for Many-Body systems}

In the context of many-body systems, the above derivations are restricted to particles interacting 
through  two-body interactions. The method presented here can also be used to provide Monte-Carlo 
formulations when particles interact through three-body or more interactions. Then higher order noises 
are necessary and evolution of the three-body or more density matrices should be estimated
along stochastic paths.     

Another possible application of the method is to include pairing correlation  
in the trial states. In that case, the relevant degrees of freedom along the path are not only the 
one-body density $\rho_{ij} = Tr(a^+_i a_jD)$ but also the abnormal density
defined as $\kappa_{ij} = Tr(a_j a_iD)$ (see for instance \cite{Rin80,Bla86}).  
Although we do not develop here explicitly the formalism, it would be interesting 
to make the connection between 
the technique described above 
and the recent related works which use either directly a stochastic simulation 
in phase-space \cite{Cor04} or stochastic Schroedinger equations \cite{Mon06,Lac06}.
   
We guess that the result is the one obtained recently in ref. \cite{Lac06}.
In that case, it has been shown that the exact two-body problem can be mapped into a quantum 
Monte-Carlo process between densities 
$D= \left| \Phi_a \right> \left< \Phi_b \right|$ written as a product of two Hartree-Fock Bogolyubov states with 
$\left< \Phi_b \left.  \right| \Phi_a \right>=1$ along the path.

\section{Discussion}
Different examples given above illustrate how one can take advantage of both 
phase-space evolution and Monte-Carlo process in Hilbert space. Most often, the exact problem
is replaced by a set of coupled non-linear stochastic equations. Therefore, we would like to mention
that we can end with same difficulties encountered sometimes in that case 
\cite{Gar00} with the appearance of unstable trajectories. This is the case for instance with the 
stochastic nonlinear equations derived in section (\ref{sect:kerr}), which are numerically
difficult to integrate \cite{Pli01}. 

Another example is given by the two-mode Hamiltonian of ref. \cite{Mil97} given 
by 
\begin{eqnarray}
H = \frac{\hbar \Omega}{2} \left( c^+_1 c_2 + c^+_2 c_1 \right) + \hbar K \left[ (c^+_1)^2 c^2_1 
+ (c^+_2)^2 c^2_2 \right], \nonumber
\end{eqnarray}
where $(c^+_1,c_1)$ and $(c^+_2,c_2)$ correspond respectively to the creation/annihilation operators 
of two orthogonal single particles states denoted by $\left| u_1 \right>$ and $\left| u_2 \right>$, and obey boson commutation 
rules. This Hamiltonian has been used to mimic the dynamics of two coupled condensates. 
The exact static and dynamical solutions associated to this Hamiltonian can be obtained exactly 
and it has been retained as a test case in ref. \cite{Car01} for Monte-Carlo methods.
The framework presented here leads to an alternative stochastic processes between densities 
$D =\left| N: \phi_a \right> \left< N: \phi_b \right|$, where single-particle states $\left| \phi_a \right>$   
and $\left|  \phi_b \right>$ can be decomposed as
\begin{eqnarray}
\left| \phi_a \right> &=& \alpha_a  (t) \left| u_1 \right> + \beta_a (t) \left| u_2 \right> \nonumber , \\
\left| \phi_b \right> &=& \alpha_b  (t) \left| u_1 \right> + \beta_b (t) \left| u_2 \right> \nonumber .
\end{eqnarray}
A direct application of the formalism developed for bosons in section \ref{sect:boson} with 
$O_1 = \sqrt{\hbar K} c^+_1c_1$ and $O_2 = \sqrt{\hbar K} c^+_2c_2$ leads to stochastic equations for 
$\alpha_a$ and $\beta_a$
\begin{widetext}
\begin{eqnarray}
d\alpha_a &=& -\frac{idt\Omega}{2} \beta_a -2idt K (N-1) \left[ \alpha^*_b \alpha_a \right] \alpha_a + 
\sqrt{2K\hbar}\left[ d\xi^{[2]}_1 -dW  \right] \alpha_a \nonumber ,\\
d\beta_a &=& -\frac{idt\Omega}{2} \alpha_a -2idt K (N-1) \left[ \beta^*_b \beta_a \right] \beta_a 
 +  \sqrt{2K\hbar} \left[ d\xi^{[2]}_2-dW  \right] \beta_a .\nonumber 
\end{eqnarray}
\end{widetext}
where $dW = d\xi^{[2]}_1 \alpha^*_b \alpha_a + d\xi^{[2]}_2 \beta^*_b \beta_a$. These equations have to be 
completed with similar equations for $\alpha^*_b$ and $\beta^*_b$. It could in particular be checked that 
$Tr(P_\phi) = \alpha^*_b \alpha_a + \beta^*_b \beta_a$ is constant along each path. 

\begin{figure}[tbph]
\includegraphics[height=8.5cm,angle=-90.]{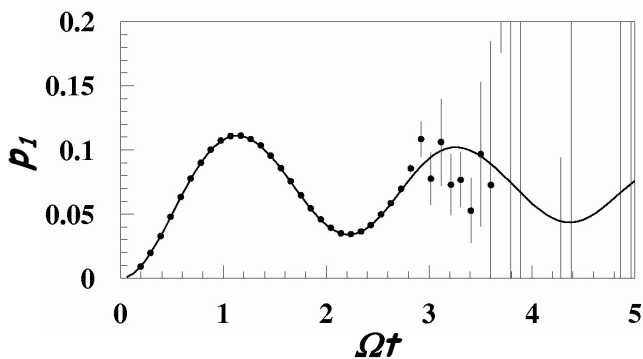}
\caption{Evolution of $p_1=\left< \right. c^+_1 c_1 \left. \right>$ as a function of time (circles) 
obtained by averaging over $10^{4}$ stochastic trajectories. The result is compared 
to the exact evolution (solid lines). Error bars represent standard deviation.}
\label{fig1:phasespace}
\end{figure}

An illustration of the result obtain by a 
direct numerical integration of these equation is shown in figure \ref{fig1:phasespace} for 
the special case where all bosons are initially in the same state $\left| u_2 \right>$
and for the same parameters values taken as in ref. \cite{Car01}, i.e. $N=17$, $K = 0.1 \Omega$ and $\Omega dt =10^3$.   
Figure \ref{fig1:phasespace} presents the evolution of $p_1 = \left< c^+_1 c_1 \right>$ as a function of 
time. 
The stochastic calculation does perfectly reproduce the exact results up to $\Omega t \simeq 2.8$
with a rather small number of trajectories. After this time, large deviations with respect 
to the exact evolution occur, and could not be reduced by increasing the number of
trajectories. A careful analysis show that some trajectories becomes hard to integrate 
numerically. Indeed, for some trajectories 
while $Tr(P_\phi) =1$, $\left| \left| \alpha^*_b \alpha_a \right| \right|$ 
and $\left| \left| \beta^*_b \beta_a \right| \right|$ may become very large. 
Accordingly, the implementation of the above stochastic equations requires 
the time step to be very small or, alternatively to develop specific numerical techniques.

This last example illustrates that the stochastic equations  
might be difficult to integrate numerically due to their non-linearity. This is a problem which seems to be recurrent
in the context of quantum stochastic mechanics both with Stochastic Schroedinger Equation \cite{Car01} or
stochastic evolution in phase-space \cite{Gar00}. Therefore, to take fully advantage of these techniques 
one should develop specific numerical methods. 
This has been done for instance in refs. \cite{Car01,Pli01,Deu01} using the fact 
that stochastic equations are generally not unique. Such a non-uniqueness also exists in 
the formalism described here.       
  
\section{Conclusion}

The main goal of this paper is to show that one can take advantage
of the phase-space evolution to construct Monte-Carlo processes in Hilbert 
space of a restricted class of trial wave-functions. In the first part 
of this work, we show that given a class of trial wave-functions $\left| Q_a \right>$ 
associated to a set of specific operators $A_\alpha$, it is always possible to obtain 
a hierarchy of stochastic approximations of a quantum problem in terms of quantum jumps between 
densities formed of couples of trial states $D=\left| Q_a \right>\left< Q_b \right|$. 
At the level $k$ of the hierarchy, the existence of such a stochastic process 
is proved using variational techniques. The quantum diffusion obtained in such a way
has a clear interpretation in phase-space evolution. Indeed, at a given level $k$
we show that moments of $A_\alpha $ calculated by averaging expectation values $\left< A_{\alpha_1} \right>$, $\left< A_{\alpha_1} A_{\alpha_2} \right>$,...,
$\left< A_{\alpha_1} \cdots A_{\alpha_k} \right>$ over different trajectories match the exact evolution. 
The stochastic formulation can eventually be exact if the Hamiltonian applied to the trial state 
can be recast as a polynomial of $A_\alpha$. 

The proof of the existence of such a hierarchy of stochastic formulation 
is very helpful to bridge stochastic mechanics in Hilbert space and phase-space 
evolution. In the second part of this article, several examples illustrates how 
the Ehrenfest theorem can directly be used to guess stochastic Schroedinger 
equations. Finally, a critical discussion on numerical aspects is given.

{\bf ACKNOWLEDGMENTS}

The author is grateful to Thomas Duguet and Vincent Rotival for
the careful reading of the manuscript and for the hospitality and financial support of the 
NSCL (Michigan State University, USA) where this work has been done in part.

\end{document}